# Visualizing *d*-wave pairing symmetry at the atomic scale in a cuprate high-$T_c$ superconductor by Josephson tunneling

Zhen Liang[1,2,3,#], Tianheng Wei[1,2,#], Jinpeng Lu[1,2,#], Wei Ren[1,2], Takenori Fujii[4], Takao Watanabe[5,6,7], Ziqiang Wang[8] & Jian Wang[1,2,3,9,*]

[1]International Center for Quantum Materials, School of Physics, Peking University, Beijing 100871, China

[2]Beijing Key Laboratory of Quantum Devices, Peking University, Beijing 100871, China

[3]Hefei National Laboratory, Hefei 230088, China

[4]Cryogenic Research Center, the University of Tokyo, Bunkyo, Tokyo 113-0032, Japan

[5]Graduate School of Science and Technology, Hirosaki University, Hirosaki, Aomori, 036-8561, Japan

[6]Physics Department, College of Engineering, Nihon University, Fukushima 963-8642, Japan

[7]Department of Advanced Materials Science, the University of Tokyo, Kashiwa, Chiba 277-8561, Japan

[8]Department of Physics, Boston College, Chestnut Hill, MA 02467, USA

[9]Collaborative Innovation Center of Quantum Matter, Beijing 100871, China

[#]These authors contributed equally.

[*]Corresponding to: jianwangphysics@pku.edu.cn (J.W.).

**Superconducting pairing symmetry provides crucial insight into the microscopic mechanisms of Cooper pairing. However, probing the local manifestations of sign-changing superconducting order parameters remains experimentally challenging. Here, we investigate the superconducting condensate in optimally doped (OPT) $Bi_2Sr_2Ca_2Cu_3O_{10+\delta}$ (Bi-2223) using scanning Josephson tunneling microscopy (SJTM), equipped with an *s*-wave superconducting tip. Measurements at ultra-low temperatures reveal a pronounced modulation of the Josephson tunneling signal at different locations inside the crystallographic unit cell. A distinct Josephson tunneling signal is detected above O sites in the underlying $CuO_2$ plane, whereas the signal nearly vanishes above the Cu-plaquette diagonals, including Cu sites and plaquette centers. This unique intra-unit-cell spatial pattern agrees with the theoretical prediction for the symmetry-dependent cancellation of Josephson tunneling between an *s*-wave superconducting tip and a *d*-wave superconducting sample. Our findings provide atomic-scale, real-space evidence for *d*-wave pairing symmetry in OPT high-$T_c$ Bi-2223 and establish SJTM as a powerful approach for investigating the pairing symmetry of unconventional superconductors.**

The symmetry of the superconducting order parameter is fundamental to understanding the nature of a superconducting state. It characterizes the transformation properties of the gap function under crystal symmetry operations and distinguishes different pairing states. Conventional superconductors described by the Bardeen-Cooper-Schrieffer (BCS) theory[1] are typically characterized by s-wave pairing, with an isotropic superconducting gap that does not change sign around the Fermi surface. In contrast, unconventional superconductors can exhibit pairing states with sign-changing gap functions under crystal symmetry rotations. Cuprate high-transition-temperature (high-$T_c$) superconductors are widely considered to exhibit predominantly *d*-wave pairing symmetry[2–4], with the superconducting order parameter changing sign under a π/2 rotation and vanishing along the nodal directions. The *d*-wave pairing state in cuprates hase been demonstrated by several phase-sensitive Josephson junction experiments[4–6]. However, finite *c*-axis Josephson tunneling currents have been observed in Pb/cuprate junctions[7–9], which is inconsistent with the cancellation of Josephson coupling expected for an ideal *s*-wave/*d*-wave junction. Moreover, twist-junction experiments have raised questions about possible *s*-wave pairing components in cuprates[10–12]. Because spatial inhomogeneities and interface effects can complicate the interpretation of macroscopic and mesoscopic Josephson junction measurements[2], directly visualizing the pairing symmetry of cuprate high-$T_c$ superconductors at the microscopic atomic scale is highly desirable.

Scanning Josephson tunneling microscopy (SJTM) provides access to local Cooper-pair tunneling through a junction formed between a superconducting tip and the superconducting sample, enabling spatially resolved atomic scale investigations of the superconducting condensate[13–22]. A recent work has demonstrated the capability of SJTM for resolving pairing symmetry at the atomic scale in real space[23]. Here, by performing SJTM using an *s*-wave Pb-coated superconducting tip, we resolved pronounced variations of local Josephson tunneling within the crystallographic unit cell on an optimally doped (OPT) $Bi_2Sr_2Ca_2Cu_3O_{10+\delta}$ (Bi-2223) single crystal, which has the highest $T_c$ among $Bi_2Sr_2Ca_{n-1}Cu_nO_{2n+4+\delta}$ (BSCCO) family compounds[24]. Distinct Josephson tunneling signals at be detected above O sites in the underlying $CuO_2$ plane, whereas the signals are strongly suppressed above the Cu-plaquette diagonals (connecting next-nearest-neighbor Cu sites), including Cu sites and plaquette center hollow sites. The observed site dependence of Josephson tunneling agrees with the theoretical predictions[25], in which *d*-wave pairing imposes symmetry-enforced cancellation of the Josephson coupling along the diagonal lines of the Cu-plaquette. Our findings provide atomic-scale, real-space evidence for *d*-wave pairing symmetry in Bi-2223 and demonstrate the capability of SJTM with an *s*-wave superconducting tip to probe the pairing symmetry of unconventional superconductors at the atomic scale.

**Characterization of the Bi-2223 sample**

The Bi-2223 single crystal was cleaved at 80 K under ultra-high vacuum, and then promptly transferred into the STM chamber at liquid-helium temperature. The Bi-2223 crystal structure comprises a BiO-SrO-$CuO_2$-Ca-$CuO_2$-Ca-$CuO_2$-SrO-BiO sequence, as schematically shown in Fig. 1a. The Bi-2223 crystal was cleaved between adjacent weakly bonded BiO layers. Figure 1b displays a representative STM topographic image of the cleaved BiO surface, resolving the approximately square Bi-lattice and the structural supermodulation. The supermodulation has a period of around 2.6 nm and slightly distorts the lattice structure. Figure 1c displays a typical differential conductance (d$I$/d$V$) tunneling spectrum acquired with a metallic PtIr tip at 40 mK. The tunneling spectrum exhibits the two-gap feature, consistent with previous studies on OPT Bi-2223[26,27]. The superconducting gap energies can be extracted by fitting the coherence peaks with double-Lorentzian functions (see Methods)[26,27]. The resulting distributions of the superconducting gap magnitudes ($\Delta_1$ and $\Delta_2$) are statistically summarized in Fig. 1d, with the average values of $\Delta_1$ and $\Delta_2$ being 43.8 meV and 52.8 meV, respectively.

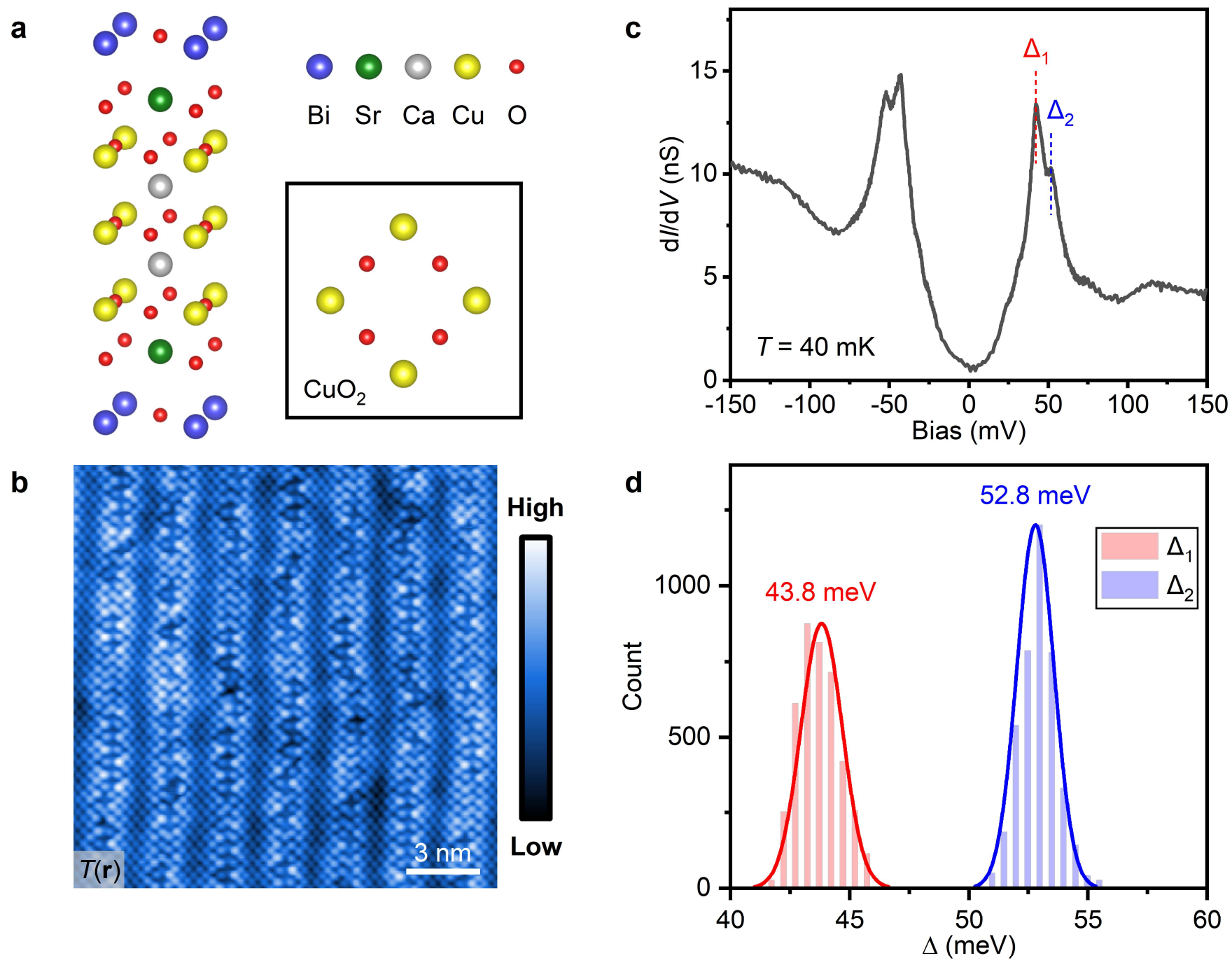

**Fig. 1 | STM characterization of the cleaved Bi-2223 single crystal. a**, The schematic of the Bi-2223 crystal structure (left) and the $CuO_2$ plane (right). **b**, A representative STM topographic image of the exposed BiO surface (17.14 × 17.14 nm$^2$, $V_{set}$ = 1 V, $I_{set}$ = 100 pA), clearly resolving the Bi-lattice and structural supermodulation with a period of around 2.6 nm. **c**, Typical d$I$/d$V$ tunneling spectrum acquired with a metallic PtIr tip ($V_{set}$ = 150 mV, $I_{set}$ = 1 nA, $V_{mod}$ = 1 mV, $T$ = 40 mK), showing two sets of superconducting coherence peaks assigned to $\Delta_1$ and $\Delta_2$. **d**, Distribution of $\Delta_1$ (red) and $\Delta_2$ (blue) with Gaussian fits. The average values of $\Delta_1$ and $\Delta_2$ are 43.8 meV and 52.8 meV, respectively.

## Josephson tunneling signal detected on the Bi-2223 sample

To further investigate the local superconducting pairing in the Bi-2223 crystal, we performed SJTM experiments using an $s$-wave superconducting tip at ultra-low temperatures (below 110 mK). The superconducting tip was prepared by depositing Pb atoms onto a PtIr tip using molecular beam epitaxy (MBE), as schematically shown in Fig. 2a. We calibrated the superconducting tip on a Pb(111) thin film grown on Si(111) substrate (Extended Data Figs. 1-2), demonstrating the very high quality of the superconducting tip. Then we performed SJTM experiments on the Bi-2223 crystal. Figure 2c displays a series of d$I$/d$V$ tunneling spectra acquired under different tip-sample distances at the position marked by the red dot in Fig. 2b. The spectra clearly show sharp coherence peaks arising from the superconducting tip, with the gap magnitude of $\Delta_{tip}$ = 1.41 meV.

Under our experimental setup conditions, the tip-vacuum-sample junction operated in the phase-diffusive regime, in which the Josephson current exhibits extrema at nonzero bias voltages[13–15,28–30]. Figure 2d shows the zoom-in d$I$/d$V$ tunneling spectra near the Fermi level, and the simultaneously acquired current-voltage ($I$-$V$) curves are presented in Fig. 2e. The zero-bias conductance ($g_J$) and the local maximum current ($I_m$) originating from Josephson tunneling can be observed, which can provide information about local superconductivity. $I_m$ is proportional to $g_J$ (Fig. 2f), and both quantities are proportional to the square of the Josephson critical current ($I_c$)[13–15,28–30].

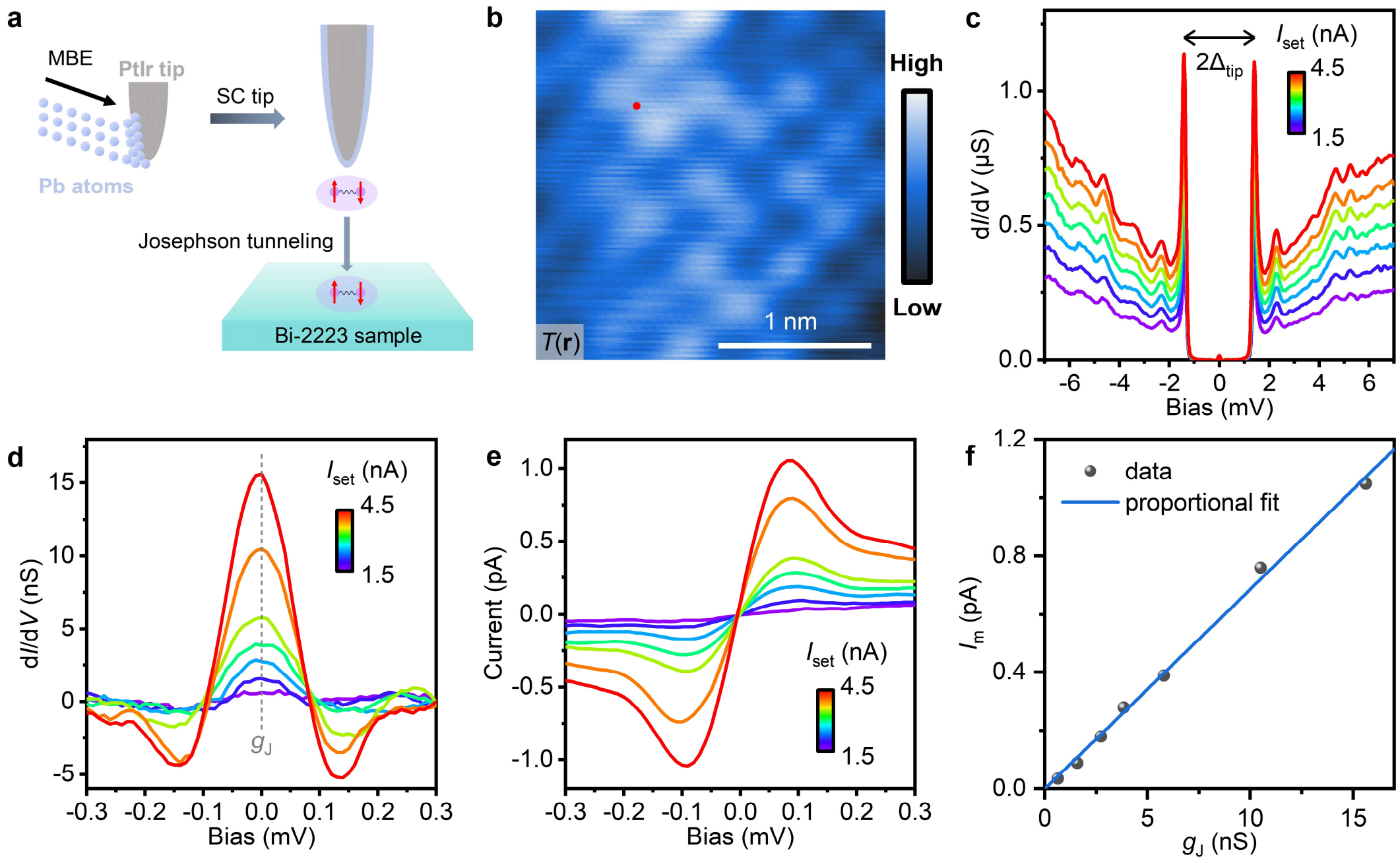

**Fig. 2 | Josephson tunneling spectra on the cleaved Bi-2223 crystal.** **a**, A schematic of the SJTM setup with an $s$-wave superconducting tip. **b**, An atomically resolved STM topographic image of the cleaved Bi-2223 crystal (2.29 × 2.29 nm$^2$, $V_{set}$ = -90 mV, $I_{set}$ = 3 nA). **c**, d$I$/d$V$ tunneling spectra acquired under different tip-sample distances at the position marked by the red dot in **b** ($V_{set}$ = -7 mV, $V_{mod}$ = 30 μV, $T$ = 85 mK, the tip approaches the sample as the setpoint current $I_{set}$ increases). Sharp coherence peaks associated with the superconducting tip yield $\Delta_{tip}$ = 1.41 meV. **d**, Zoom-in of the spectra around zero bias of **c**, showing the zero-bias conductance $g_J$ due to Josephson tunneling. **e**, $I$-$V$ curves simultaneously acquired with **d**. **f**, Relationship between the local maximum current $I_m$ and $g_J$, which can be fitted well by a proportional function. $I_m$ is extracted as half of the difference between the tunneling current at ±90 μV for each $I$-$V$ curve in **e**.

## Site dependence of the Josephson tunneling

To reveal the site dependence of the Josephson tunneling signal, we measured the Josephson tunneling spectra above distinct atomic sites of the underlying $CuO_2$ plane, as shown in Fig. 3. The Cu sites project onto the surface Bi sites, whereas the O sites project onto the midpoints between nearest-neighbor Bi atoms. The hollow sites correspond to the centers of the Cu-plaquettes (midpoints between next-nearest-neighbor Cu atoms) and, in projection, the centers of the surface Bi-plaquettes (midpoints between next-nearest-neighbor Bi atoms). Distinct Josephson tunneling signals are detected above the O sites of the $CuO_2$ plane. Strikingly, above the Cu sites and hollow sites of the $CuO_2$ plane, the Josephson tunneling signal almost vanishes. This contrast is evident in both the individual spectra (Fig. 3b-c) and their site-resolved averages (Fig. 3d-e). The same site dependence can be reproduced in different sample regions (Extended Data Figs. 3-4), demonstrating that our observations are reliable.

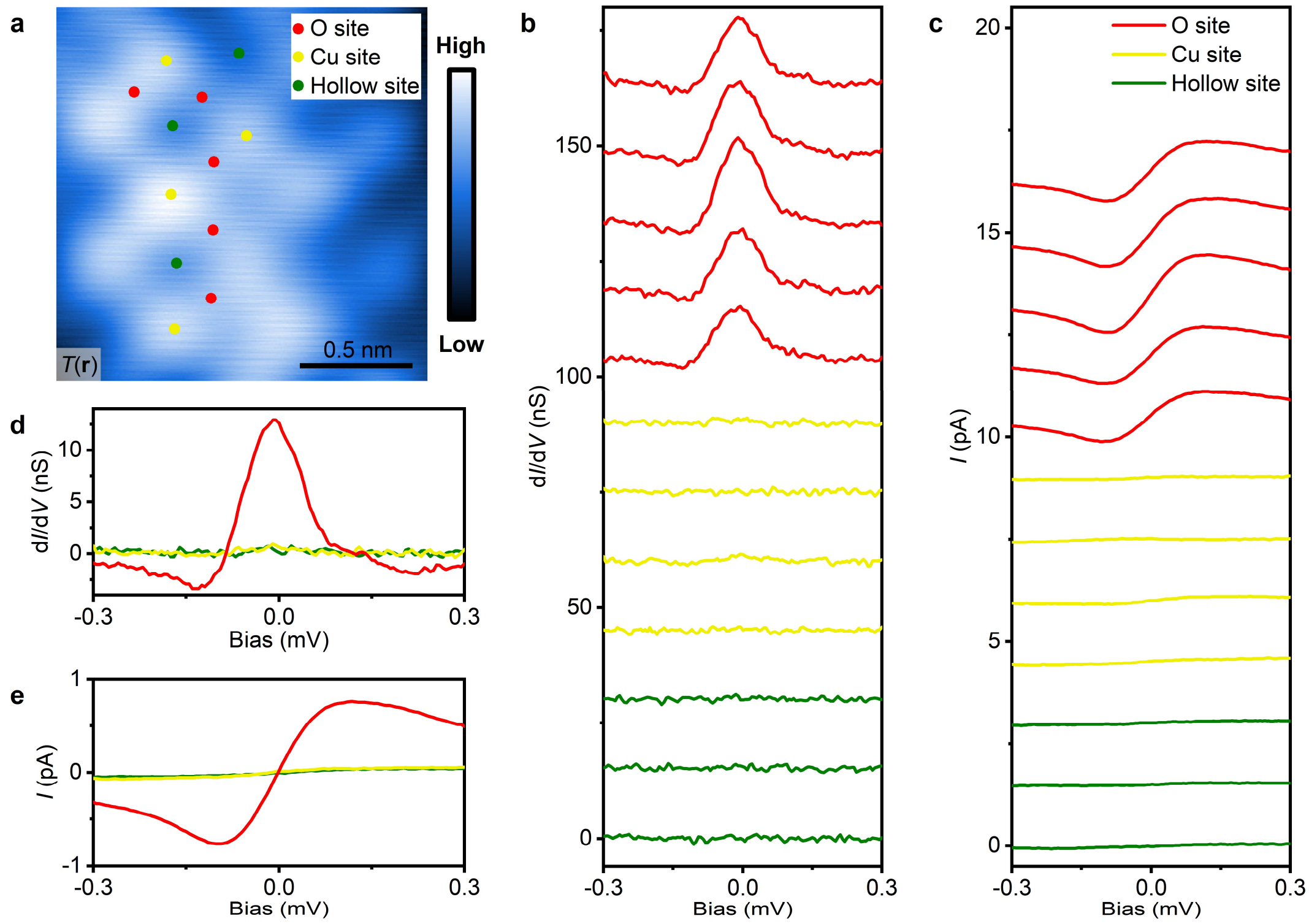

**Fig. 3 | Site-dependent Josephson tunneling on the Bi-2223 crystal. a**, An atomically resolved STM topographic image of the cleaved Bi-2223 crystal (1.64 × 1.64 nm$^2$, $V_{set}$ = -90 mV, $I_{set}$ = 100 pA). The colored solid circles mark the projected location of the underlying O atom, Cu atom, and hollow sites superimposed on the top of the topographic image. **b**, d$I$/d$V$ tunneling spectra acquired above O sites (red curves), Cu sites (yellow curves) and hollow sites (green curves) of the underlying $CuO_2$ plane ($V_{set}$ = -7 mV, $I_{set}$ = 4 nA, $V_{mod}$ = 50 μV, $T$ = 94 mK). The spectra are vertically shifted for clarity. The corresponding measurement positions are marked by the colored solid circles in **a**. **c**, $I$-$V$ curves simultaneously acquired with **b**. The curves are vertically shifted for clarity. **d**, Site-resolved averages of the spectra shown in **b**. **e**, Site-resolved averages of the $I$-$V$ curves shown in **c**. Distinct Josephson tunneling signals are observed above the O sites, whereas the Josephson tunneling signal almost vanishes above the Cu sites and hollow sites.

To further investigate the spatial distribution of the Josephson tunneling signal, we performed spectroscopic imaging SJTM measurements. Figure 4a shows an STM topographic image of the cleaved Bi-2223 crystal. Simultaneously, we measured the maps of the Josephson tunneling current at bias voltages of 90 μV and -90 μV in this area, as shown in Fig. 4b-c. We also plot the difference between them, as shown in Fig. 4d. The Cu sites of the underlying $CuO_2$ plane, which correspond to the surface Bi sites, are marked by yellow balls. The Josephson tunneling current maps show clear Josephson tunneling current above midpoints between nearest-neighbor Cu sites, corresponding to the O sites of the $CuO_2$ plane. In contrast, the Josephson tunneling current almost vanishes above Cu-plaquette diagonals (connecting next-nearest-neighbor Cu sites), including Cu sites and hollow sites.

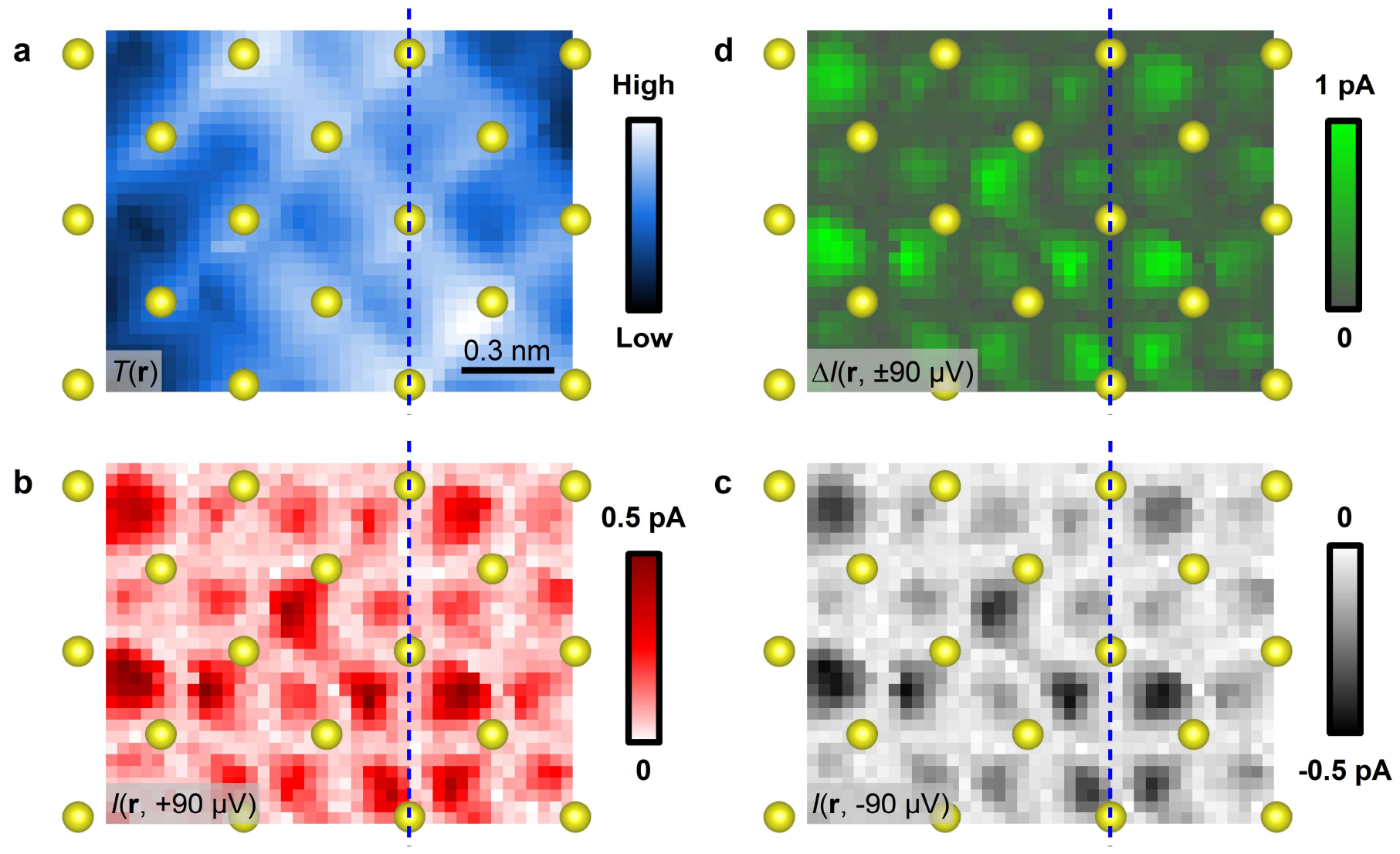

**Fig. 4 | Spatial distribution of the Josephson tunneling current. a**, An STM topographic image of the cleaved Bi-2223 crystal (1.51 × 1.17 nm$^2$, $V_{set}$ = 7 mV, $I_{set}$ = 4 nA). **b-c**, The maps of the Josephson tunneling current measured in the same area as in **a**, at bias voltages of 90 μV and -90 μV, respectively ($V_{set}$ = 7 mV, $I_{set}$ = 4 nA, $V_{mod}$ = 50 μV, $T$ = 66 mK, 703 pixels/nm$^2$). The Josephson current maps (**b-c**) and the topographic image (**a**) were obtained simultaneously, showing pixel-to-pixel correspondence between these datasets. The Cu sites (right below the surface Bi sites) are marked by yellow balls. **d**, The difference between the Josephson tunneling current measured at 90 μV and -90 μV. Clear Josephson tunneling current is detected above O sites, corresponding to the midpoints between nearest-neighbor Cu sites. In contrast, the Josephson tunneling current almost vanishes above Cu-plaquette diagonals (connecting next-nearest-neighbor Cu sites), as marked by the blue dashed lines in **a-d**.

To assess whether gap variations could account for the Josephson contrast, we acquired spatially resolved d$I$/d$V$ tunneling spectra using a metallic PtIr tip on Bi-2223 (Extended Data Fig. 5a). The superconducting gap magnitudes $\Delta_1$ and $\Delta_2$ can be extracted by fitting the coherence peaks with double-Lorentzian functions (see Methods)[26,27], as shown in Extended Data Fig. 5b-c. Both $\Delta_1$ and $\Delta_2$ only show small spatial variations across the measured area. The overall gap variations can be estimated from the standard deviations of the $\Delta_1$ and $\Delta_2$ maps, which are calculated to be 0.869 meV and 0.780 meV, respectively, corresponding to only 1.98% and 1.48% of their mean gap magnitudes. Furthermore, we also compared the averaged tunneling spectra acquired above O, Cu and hollow sites, as shown in Extended Data Fig. 6. The site-resolved averaged tunneling spectra exhibit very similar line shapes. Together, these experimental results argue against a simple explanation of the pronounced Josephson contrast based on site-dependent superconducting gap magnitudes.

**Discussion**

The remarkable site dependence of the Josephson tunneling signal provides microscopic evidence for $d$-wave pairing symmetry in the $CuO_2$ plane. In particular, the distinct Josephson tunneling above O sites, taken together with its strong suppression above Cu sites and plaquette center hollow sites, is consistent with the symmetry-dependent cancellation of the Josephson tunneling predicted for an $s$-wave superconducting tip and a $d$-wave superconducting sample[25]. Following ref.[25], the spatially dependent Josephson critical current can be expressed as

$$I_c(\boldsymbol{r}) \propto \sum_{\boldsymbol{k}\in \mathrm{BZ}} \Delta^*(\boldsymbol{k})\gamma(\boldsymbol{k})|W_{\boldsymbol{k}}(\boldsymbol{r})|^2,$$

where $\Delta(\boldsymbol{k})$ is the $d$-wave superconducting gap function of the sample, $W_{\boldsymbol{k}}(\boldsymbol{r}) = \sum_{\boldsymbol{R}} e^{i\boldsymbol{k}\cdot\boldsymbol{R}} w_{\boldsymbol{R}}(\boldsymbol{r})$ is the Fourier transform of the $\mathrm{Cu}-d_{x^2-y^2}$ Wannier function $w_{\boldsymbol{R}}(\boldsymbol{r})$, and the summation extends over the Brillouin zone. The function $\gamma(\boldsymbol{k})$ is given by

$$\gamma(\boldsymbol{k}) = \int \frac{d\omega}{2\pi} \cdot n_{\mathrm{F}}(\omega) \cdot \mathrm{Im}\left\{\frac{-i \cdot \mathrm{sgn}(\omega) \cdot \pi \cdot N_0 \cdot \Delta_{\mathrm{tip}}}{[(\omega + i0^+)^2 - E_{\boldsymbol{k}}^2] \cdot \sqrt{(\omega + i0^+)^2 - \Delta_{\mathrm{tip}}^2}}\right\}$$

where $n_{\mathrm{F}}(\omega)$ is the Fermi-Dirac distribution function, $N_0$ the Fermi surface density of states of the tip, and $E_{\boldsymbol{k}}$ the quasi-particle energy dispersion of the sample.

For $d$-wave pairing in the cuprates, the gap function changes sign under a $\pi/2$ rotation or a diagonal mirror reflection (Fig. 5a), whereas both $\gamma(\boldsymbol{k})$ and $|W_{\boldsymbol{k}}(\boldsymbol{r})|^2$ remain invariant under these symmetry operations[25,31]. These leads to

$$I_c(\hat{C}_4\boldsymbol{r}) = -I_c(\boldsymbol{r}), \qquad I_c(\hat{\sigma}\boldsymbol{r}) = -I_c(\boldsymbol{r}),$$

where $\hat{C}_4$ denotes a $\pi/2$ rotation around a Cu site and $\hat{\sigma}$ denotes a diagonal mirror reflection. For positions $\boldsymbol{r_d}$ lying on a diagonal mirror line, $\hat{\sigma}\boldsymbol{r_d} = \boldsymbol{r_d}$, the symmetry transformation requires $I_c(\boldsymbol{r_d}) = 0$. Thus, the Josephson critical current vanishes at positions directly above the Cu-plaquette diagonals, including the Cu and hollow sites. The Wannier-based calculation further predicts that $|I_c(\boldsymbol{r})|$ reaches its largest magnitude above O sites[25]. Such spatial modulation of the Josephson tunneling signal can be better understood by the schematic of the spatial distribution of $I_c(\boldsymbol{r})$ shown in Fig. 5b. Since the experimentally detected Josephson tunneling signal is proportional to the square of the Josephson critical current, $I_c^2(\boldsymbol{r})$, we plot the corresponding spatial distribution of $I_c^2(\boldsymbol{r})$ in Fig. 5c. The resulting pattern exhibits maxima above O sites and vanishing signals along the Cu-plaquette diagonals, in agreement with our SJTM observations.

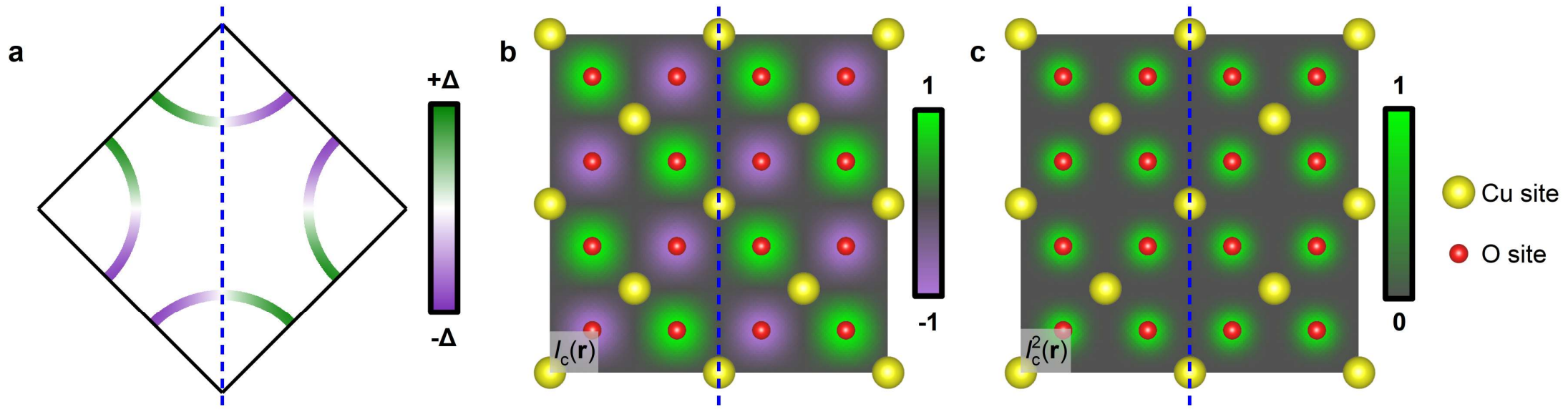

**Fig. 5 | Schematic for the spatial dependence of the Josephson coupling. a**, Schematic for *d*-wave pairing in cuprates. The gap function changes sign under a π/2 rotation or refection across a diagonal mirror plane, as indicated by the blue dashed line. **b**, Schematic for the distribution of Josephson critical current ($I_c$) between an isotropic *s*-wave superconducting tip and a *d*-wave superconducting sample. The $I_c(\mathbf{r})$ map is normalized with respect to the absolute value of $I_c$ above O sites of the $CuO_2$ plane. The blue dashed line marks a Cu-plaquette diagonal mirror line, across which $I_c(\mathbf{r})$ changes sign. **c**, The corresponding normalized $I_c^2(\boldsymbol{r})$ distribution, exhibiting the maxima above O sites and vanishing lines along the Cu-plaquette diagonals.

In conclusion, we have used SJTM with an *s*-wave superconducting tip to investigate the local Josephson tunneling response of OPT Bi-2223 single crystal at the atomic scale. We observed a pronounced intra-unit-cell modulation of the Josephson tunneling signal, which is maximized above the O sites of the $CuO_2$ plane but nearly vanishes along the Cu-plaquette diagonals, including the Cu sites and the plaquette centers. This spatial pattern agrees with theoretical predictions for Josephson tunneling between an *s*-wave superconducting tip and the *d*-wave cuprate superconductor, including the symmetry-enforced suppression along the Cu-plaquette diagonals. Our findings provide atomic-scale, real-space evidence for the *d*-wave pairing symmetry in OPT Bi-2223 single crystals and demonstrate that SJTM with an *s*-wave superconducting tip is a powerful tool to probe the pairing symmetry of unconventional superconductors.

## Methods

### STM/S measurements

The STM/S experiments were performed on a dilution refrigeration USM-1600 ultra-high vacuum MBE-STM combined system. The d$I$/d$V$ tunneling spectra were acquired by standard lock-in technique at 983 Hz.

### Double-Lorentzian fitting for the d$I$/d$V$ tunneling spectra with two-gap feature

The superconducting gap magnitudes were extracted by fitting the coherence peaks in each tunneling spectrum, following the fitting method described below[26,27]. For the positive-bias analysis, the background was estimated by fitting the data outside the 30 mV to 65 mV interval with a third-order polynomial. The fitted background was then subtracted from the measured spectrum. The resulting spectrum within the 30 mV to 65 mV interval was fitted with a double-Lorentzian function. The same procedure was applied at negative bias, with the background estimated using a third-order polynomial fitted to data outside the -65 mV to -30 mV interval. After subtracting the background, we fitted the resulting spectrum within the -65 mV to -30 mV interval with a double-Lorentzian function.

The double-Lorentzian fitting functions for the positive-bias and negative-bias spectra are given by $f^{\pm}(V) = \frac{A_1^{\pm}}{(V-\Delta_1^{\pm})^2+(K_1^{\pm})^2} + \frac{A_2^{\pm}}{(V-\Delta_2^{\pm})^2+(K_2^{\pm})^2}$. Here, the superscripts $+$ and $-$ label the positive-bias and negative-bias fits, respectively. For each component $p \in \{1, 2\}$, $\Delta_p^{\pm}$ denotes the fitting peak position. The gap magnitudes were calculated from half the energy separation between the corresponding positive-bias and negative-bias peaks: $\Delta_1 = \frac{e}{2}(\Delta_1^{+} - \Delta_1^{-})$ and $\Delta_2 = \frac{e}{2}(\Delta_2^{+} - \Delta_2^{-})$.

## Data Availability

All data that support the findings of this study are available from the corresponding author on reasonable request.

## Funding

This work was financially supported by the National Natural Science Foundation of China (Grant No. 12488201), the National Key Research and Development Program of China (Grant No. 2025YFA1411300), and The Fundamental Research Funds for the Central Universities, Peking University. T.F. and T.Watanabe acknowledge the support of the JSPS KAKENHI (25400349, 20K03849, and 23K03317).

## Author contributions

J.W. conceived and supervised the project. T.F. and T.Watanabe grew the Bi-2223 sample. Z.L., T.Wei, J.L. and W.R. carried out the STM/S experiments. Z.L, T.Wei, J.L. and J.W. analyzed the experimental data. Z.L., T.Wei, Z.W. and J.W. wrote the manuscript.

## Competing interests

The authors declare no competing interests.

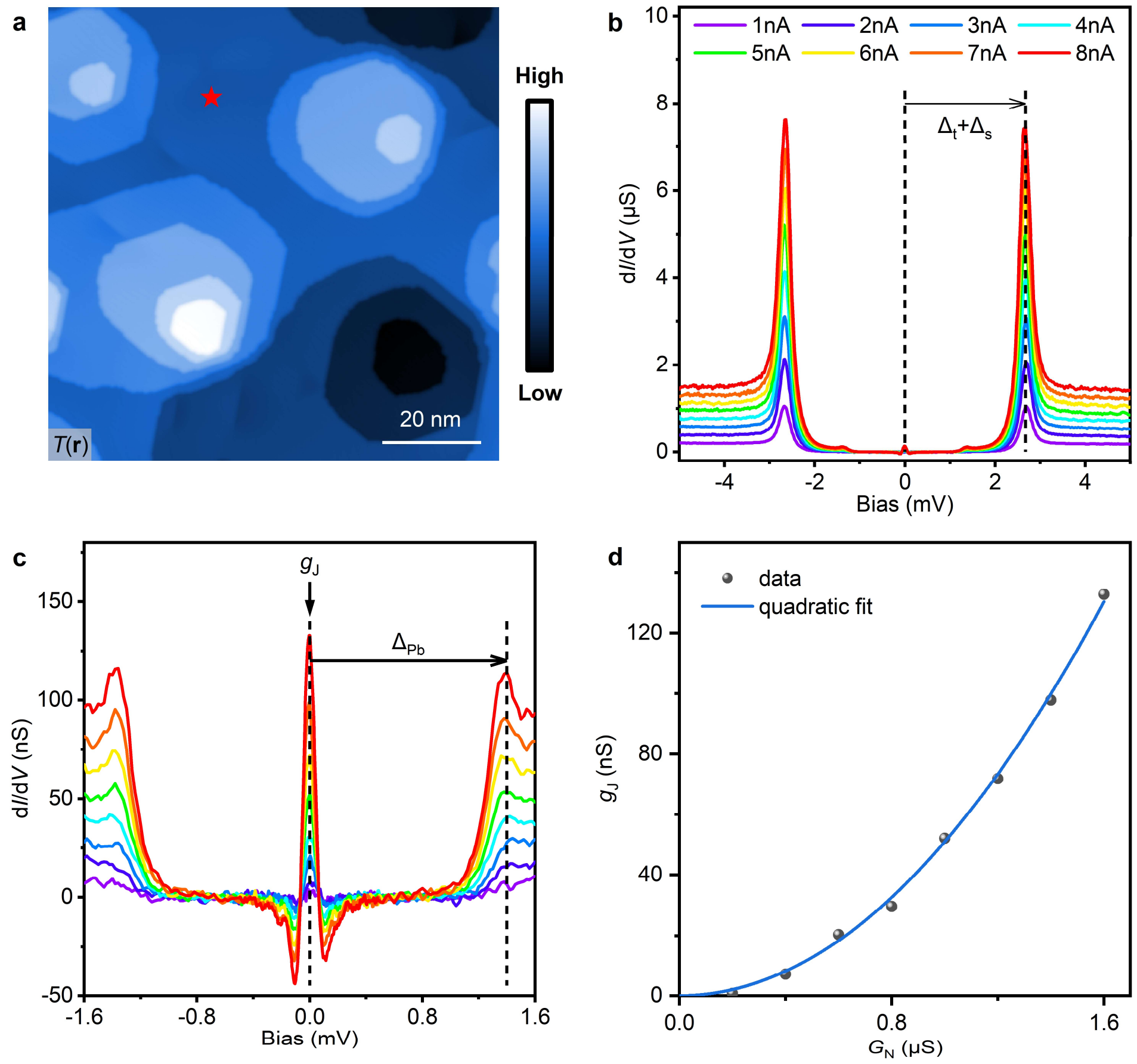

**Extended Data Fig. 1 | Josephson tunneling spectra on the Pb film. a**, A topographic image of the Pb/Si(111) film (91.43 × 91.43 nm$^2$, $V_{set}$ = 1 V, $I_{set}$ = 100 pA). Several Pb nanoislands with clear layered structure are formed on top of the Pb film. **b**, Tunneling spectra measured by a Pb-coated superconducting tip for different tip-sample distances at the red star shown in **a** ($V_{set}$ = 5 mV, $V_{mod}$ = 30 μV, $T$ = 50 mK, the tip approaches the sample as the setpoint current $I_{set}$ increases). **c**, Zoom-in spectra around zero bias of **b**. The conductance peaks at zero bias ($g_J$) and $\Delta_{Pb}$ correspond to Josephson tunneling and Andreev reflection, respectively. **d**, Quadratic fit of the Josephson tunneling conductance at zero bias $g_J$ versus the normal state conductance $G_N \equiv I_{set}/V_{set}$, which is a characteristic of Josephson effect.

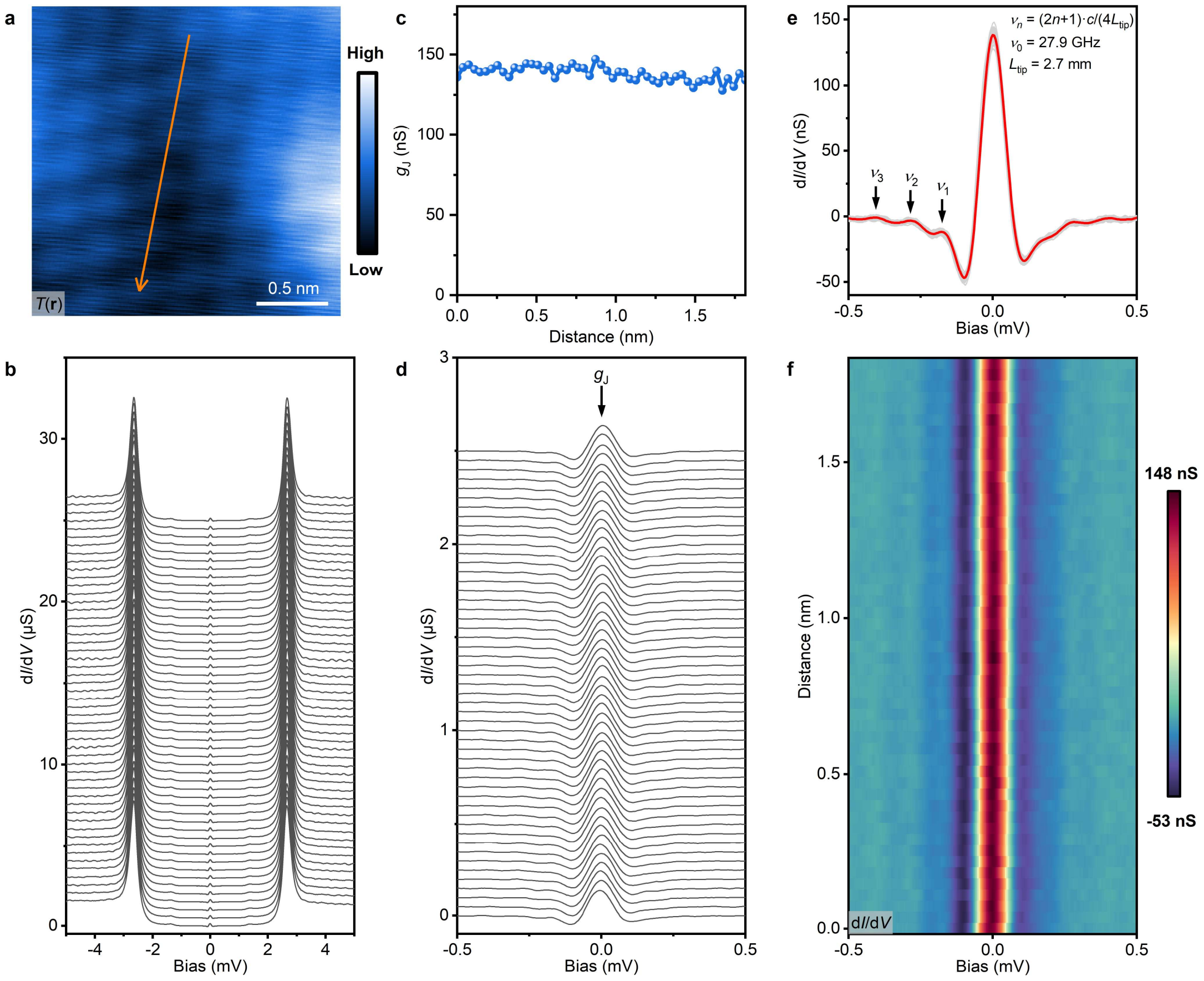

**Extended Data Fig. 2 | Josephson tunneling spectra measured along a line profile on the Pb film. a**, A topographic image with atomic resolution of the Pb/Si(111) film ($2.15 \times 2.15$ nm$^2$, $V_{set}$ = 5 mV, $I_{set}$ = 8 nA). **b**, Tunneling spectra measured along the orange arrow shown in **a** ($V_{set}$ = 5 mV, $I_{set}$ = 8 nA, $V_{mod}$ = 30 μV, $T$ = 50 mK). The spectra are vertically shifted for clarity. **c**, Josephson tunneling conductance at zero bias ($g_J$) along the orange arrow in **a**. **d**, Zoom-in spectra around zero bias of **b**. The spectra are vertically shifted for clarity. **e**, The red curve is the averaged tunneling spectrum around zero bias. The peaks marked by $\nu_{1,2,3}$ correspond to photon-assisted inelastic Cooper pair tunneling with a characteristic frequency of $\nu_0$ = 27.9 GHz, consistent with the eigenmodes of a 2.7-mm-long tip used in our SJTM studies[32]. **f**, The color plot of **d**.

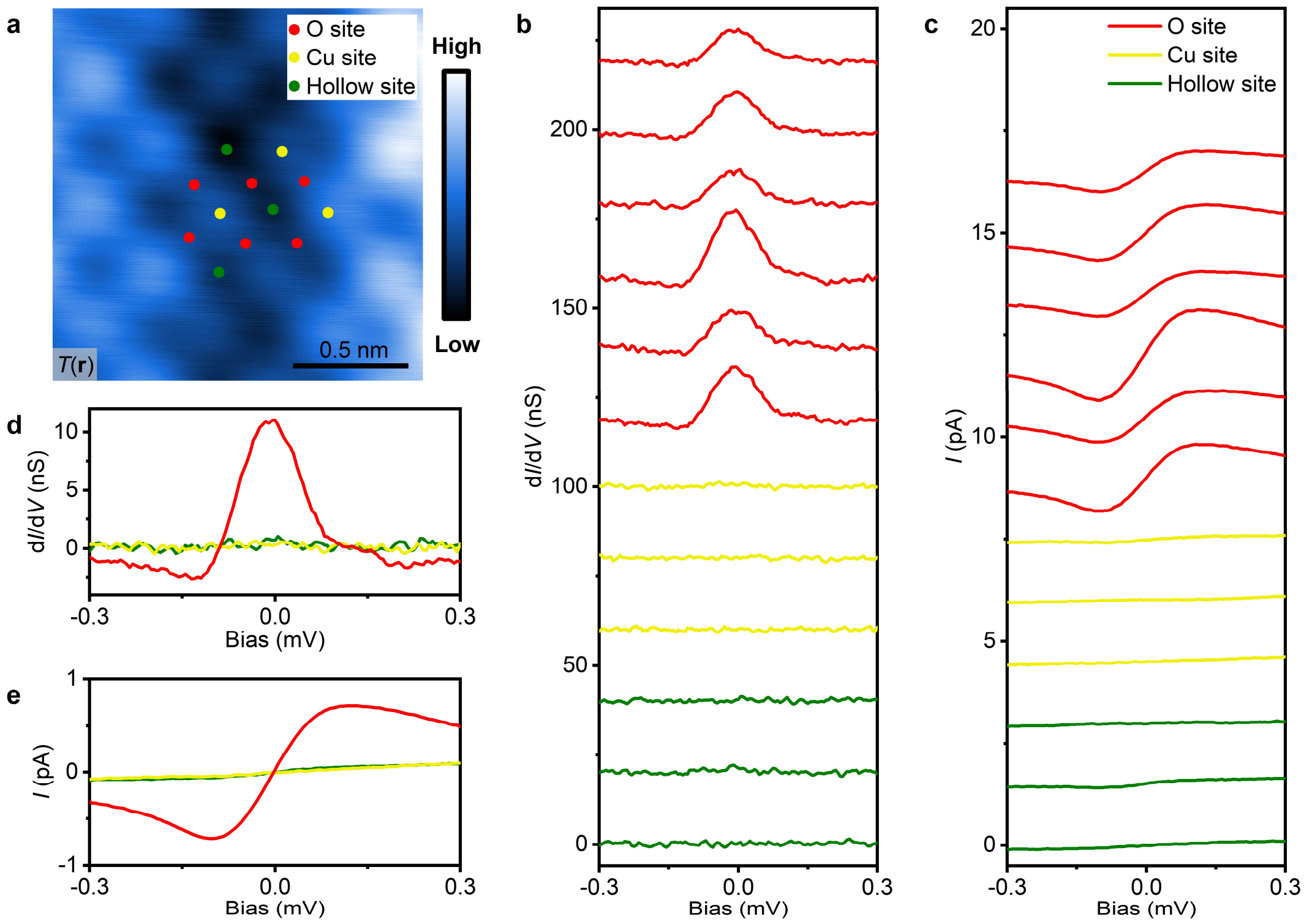

**Extended Data Fig. 3 | Josephson tunneling spectra measured at different positions.** **a**, An atomically resolved STM topographic image of the Bi-2223 crystal ($1.61 \times 1.61$ nm$^2$, $V_{set}$ = -90 mV, $I_{set}$ = 100 pA). The colored solid circles mark the projected location of the underlying O atom, Cu atom, and hollow sites superimposed on the top of the topographic image. **b**, d$I$/d$V$ tunneling spectra acquired above O sites (red curves), Cu sites (yellow curves) and hollow sites (green curves) of the underlying $CuO_2$ plane ($V_{set}$ = -7 mV, $I_{set}$ = 4 nA, $V_{mod}$ = 50 µV, $T$ = 85 mK). The spectra are vertically shifted for clarity. The corresponding measurement positions are marked by the colored solid circles in **a**. **c**, $I$-$V$ curves simultaneously acquired with **b**. The curves are vertically shifted for clarity. **d**, Site-resolved averages of the spectra shown in **b**. **e**, Site-resolved averages of the $I$-$V$ curves shown in **c**. Distinct Josephson tunneling signals are observed above the O sites, whereas the Josephson tunneling signal almost vanishes above the Cu sites and hollow sites.

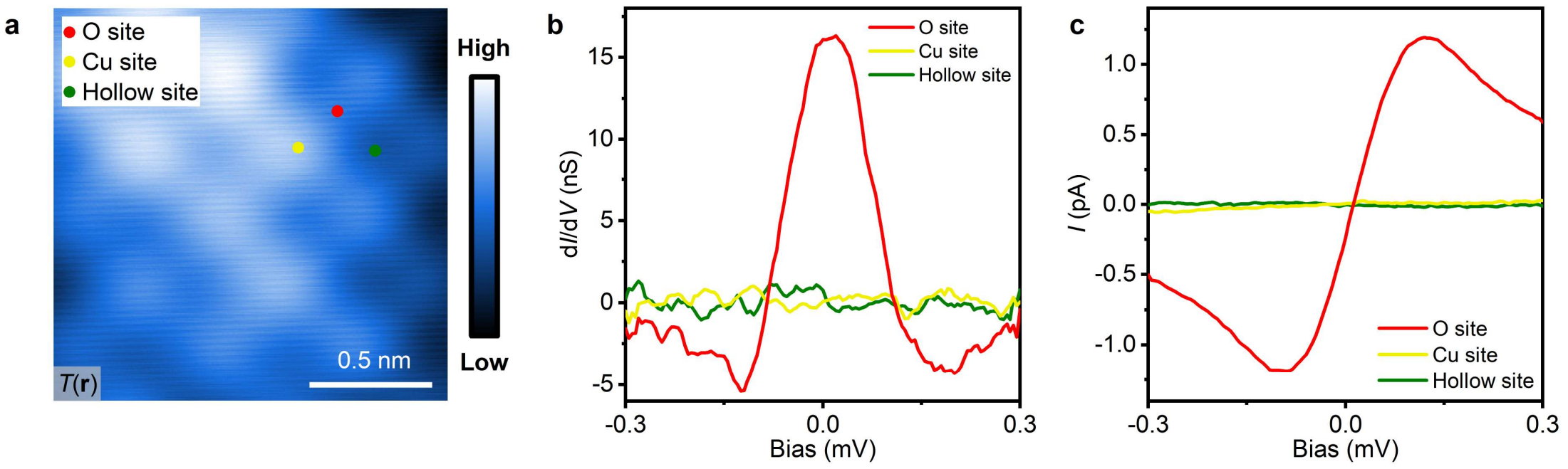

**Extended Data Fig. 4 | Josephson tunneling spectra measured at different positions.** **a**, An atomically resolved STM topographic image of the Bi-2223 crystal ($1.61 \times 1.61$ nm$^2$, $V_{set}$ = -90 mV, $I_{set}$ = 100 pA). The colored solid circles mark the projected location of the underlying O atom, Cu atom, and hollow sites superimposed on the top of the topographic image. **b**, $dI/dV$ tunneling spectra acquired above the O site (red curve), Cu site (yellow curve) and hollow site (green curve) of the underlying $CuO_2$ plane ($V_{set}$ = -7 mV, $I_{set}$ = 4 nA, $V_{mod}$ = 50 μV, $T$ = 106 mK). The corresponding measurement positions are marked by the colored solid circles in **a**. **c**, $I$-$V$ curves simultaneously acquired with **b**. Clear Josephson tunneling signal is observed above the O site, whereas the signal almost vanishes above the Cu site and hollow site.

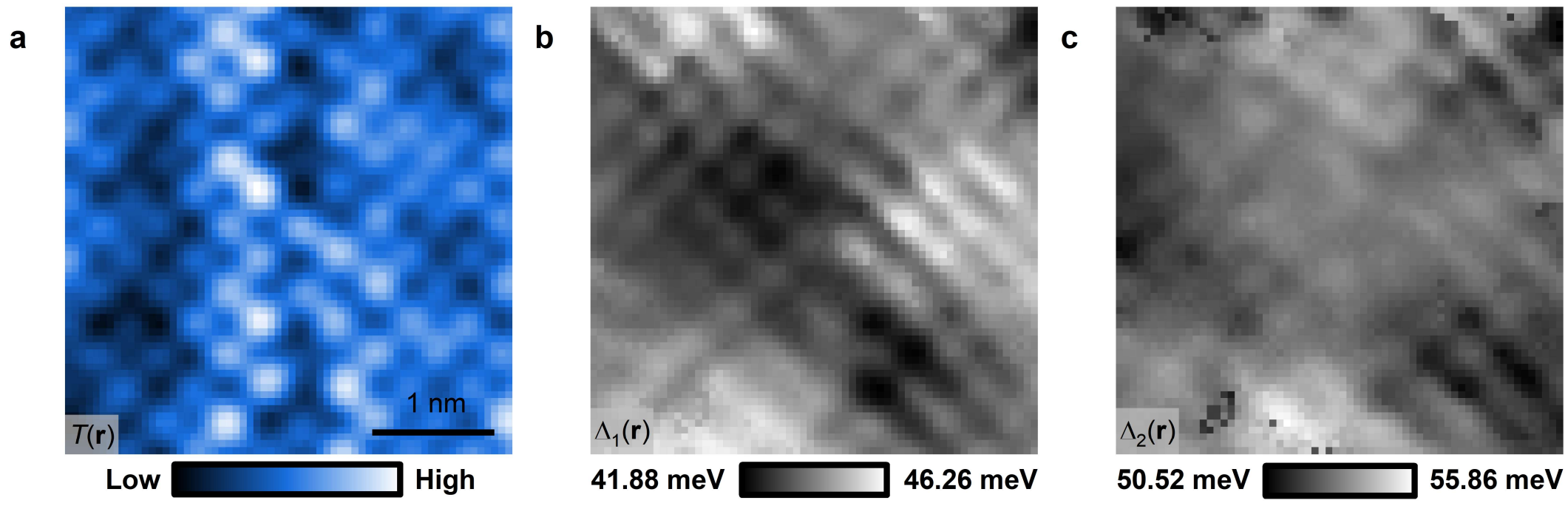

**Extended Data Fig. 5 | Spatial distribution of the superconducting gaps. a**, An atomically resolved STM topographic image of the Bi-2223 crystal (3.66 × 3.66 nm$^2$, $V_{set}$ = 150 mV, $I_{set}$ = 1 nA, 306 pixels/nm$^2$). **b-c**, Maps of $\Delta_1$ and $\Delta_2$, respectively ($V_{set}$ = 150 mV, $I_{set}$ = 1 nA, $V_{mod}$ = 1 mV, $T$ = 40 mK). The gap magnitudes were extracted by fitting the coherence peaks of the tunneling spectra (obtained using a metallic PtIr tip) with double-Lorentzian functions. The gap maps (**b-c**) and the topographic image (**a**) were obtained simultaneously, showing pixel-to-pixel correspondence between these two datasets. Both $\Delta_1$ and $\Delta_2$ show only small spatial variations across the measured area. The overall gap variations can be estimated from the standard deviations of the $\Delta_1$ and $\Delta_2$ maps, which are calculated to be only 0.869 meV and 0.780 meV, respectively, corresponding to 1.98% and 1.48% of their mean gap magnitudes.

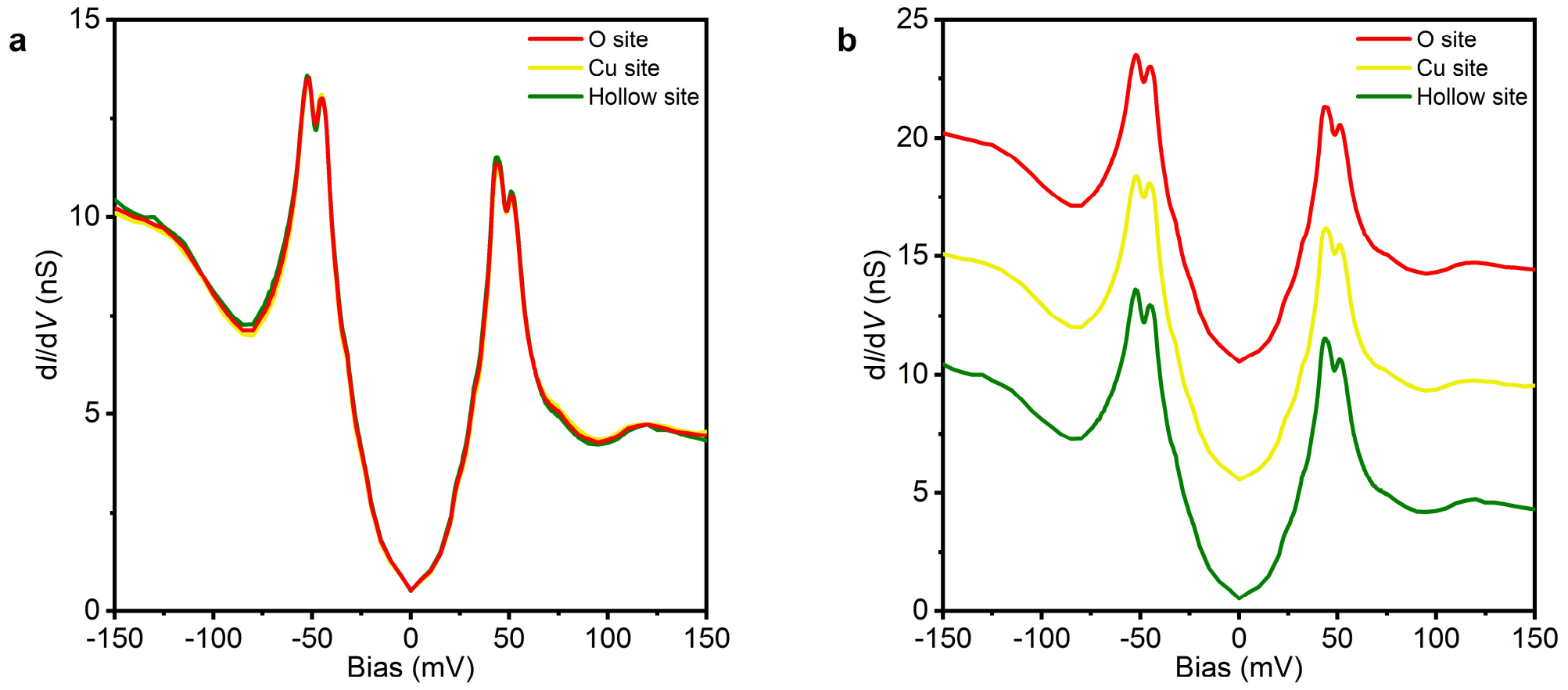

**Extended Data Fig. 6 | Spatially averaged tunneling spectra measured above different atomic sites. a**, The averaged tunneling spectra measured above different O sites (red), Cu sites (blue) and hollow sites (green) of the $CuO_2$ plane, using a metallic PtIr tip ($V_{set}$ = 150 mV, $I_{set}$ = 1 nA, $V_{mod}$ = 1 mV, $T$ = 40 mK). **b**, The same spectra as in **a** vertically shifted for clarity. The averaged spectra acquired above three types of sites exhibit very similar line shapes.